# Pseudoelasticity of SrNi$_2$P$_2$ micropillar via Double Lattice Collapse and Expansion

*Shuyang Xiao[1,*], Vladislav Borisov[2], Guilherme Gorgen-Lesseux[3], Sarshad Rommel[1], Gyuho Song[1], Jessica M. Maita[1], Mark Aindow[1], Roser Valentí[4], Paul C. Canfield[3], Seok-Woo Lee[1]*

1. Department of Materials Science and Engineering & Institute of Materials Science, University of Connecticut, 97 North Eagleville Road, Unit 3136, Storrs, CT 06269-3136, USA

2. Department of Physics and Astronomy, Uppsala University, Box 516, SE-75120 Uppsala, Sweden

3. Ames Laboratory & Department of Physics and Astronomy, Iowa State University, Ames, IA  50011, USA

4. Institute of Theoretical Physics, Goethe University, Frankfurt am Main, D-60438 Frankfurt am Main, Germany

**Abstract**

The maximum recoverable strain of most crystalline solids is less than 1% because plastic deformation or fracture usually occurs at a small strain. In this work, we show that a $SrNi_2P_2$ micropillar exhibits pseudoelasticity with a large maximum recoverable strain of ~14% under uniaxial compression via unique reversible structural transformation, double lattice collapse-expansion that is repeatable under cyclic loading. Its high yield strength (~3.8±0.5GPa) and large maximum recoverable strain bring out the ultrahigh modulus of resilience (~146±19MJ/$m^3$) a few orders of magnitude higher than that of most engineering materials. The double lattice collapse-expansion mechanism shows stress-strain behaviors similar with that of conventional shape memory alloys, such as hysteresis and thermo-mechanical actuation, even though the structural changes involved are completely different. Our work suggests that the discovery of a new class of high performance $ThCr_2Si_2$-structured materials will open new research opportunities in the field of pseudoelasticity.

**Abstract Graphic**

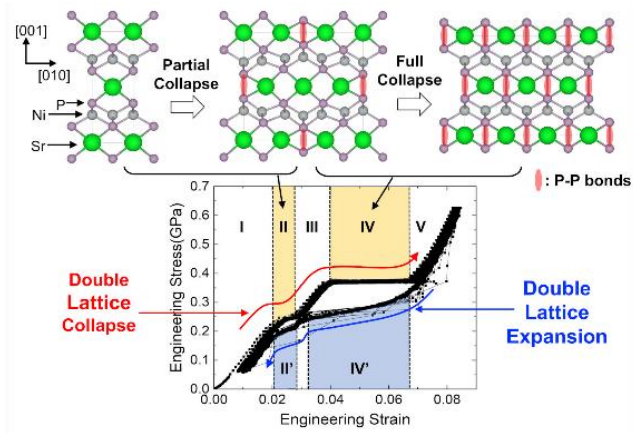

**1. Introduction**

Maximum recoverable strain, which is the measure of the maximum allowed fractional change in material length before a permanent shape change occurs, is the quantity used to describe the elastic deformability of materials.[1] The maximum recoverable strain corresponds to the safety limit for elastic shape change in an engineering design, so a high maximum recoverable strain allows a material to be used securely under a wide range of dimensional constraints.[2] Also, a high maximum recoverable strain usually allows for a large absorption of mechanical energy, providing a greater protection from yielding or failure during mechanical impact.[3,4] Furthermore, a high maximum recoverable strain enables strain engineering, which provides a pathway to control material properties even without a change in material chemistry.[5,6] Therefore, achieving a high maximum recoverable strain is certainly desirable in many aspects, but the maximum recoverable strain of most crystalline solids is, unfortunately, less than one percent. This is because crystalline defects, such as dislocations, almost always cause a permanent shape change at a small strain or at a low stress.[7,8] Therefore, it is challenging to achieve a large maximum recoverable strain in crystalline solids unless they are able to undergo a reversible structural transition.

Pseudoelastic and shape memory materials, for instance, NiTi-base or Cu-base systems, are the most widely known crystalline materials that can exhibit a large maximum recoverable strain via reversible structural transitions, twinning and de-twinning processes.[9–12] When a stimulus, such as heat or a magnetic field, is applied, the deformation-induced structural transition can be reversed, and the sample shape can be restored. If this process occurs spontaneously when an applied load is removed under the application of stimulus, it is called pseudoelasticity.[9–12] Except for only a few cases,[13–15] however, shape memory alloys accumulate permanent damage when the structural transition occurs repeatedly. The irreversible evolution of dislocation structures and the residual transformed phase impedes the shape recovery process and reduces the maximum recoverable strain under repeated loading. Thus, the improvement of structural stability has been a great concern in the field of pseudoelastic and shape memory materials.

In order to lead to a material innovation, which enables the enhancement of both the maximum recoverable strain and the structural stability, it is worthwhile to turn our attention to a new type of reversible structural transitions that has not been considered previously as a pseudoelasticity mechanism and to evaluate their pseudoelastic performance. In 1985, Hoffmann and Zhang postulated the possibility of forming and breaking Si-type bonds in $ThCr_2Si_2$-structured intermetallic compounds under uniaxial compression along their c-axis.[16] Hoffman and Zhang's postulation, as well as our recent work on $CaFe_2As_2$[17] and $CaKFe_4As_4$[18], has motivated us to perform an investigation of pseudoelastic behavior and structural stability for several $ThCr_2Si_2$-structured intermetallic compounds, single crystals of which can be grown experimentally (See also Supporting Information S1, S2, and Refs.[17–19]). By reviewing crystallographic data of a wide variety of known $ThCr_2Si_2$-structured compounds, we identified $SrNi_2P_2$ as one of the most likely candidates to have a relatively small critical stress, which is probably much lower than fracture strength, for lattice collapse.

In this study, we report that a SrNi$_2$P$_2$ single crystal, exhibits large maximum recoverable strain of 14.1±0.4% via unique structural transition, double lattice collapse-expansion under uniaxial compression along the c-axis. Note that our uniaxial mechanical results are unique, compared to the other SrNi$_2$P$_2$ works, which performed the hydrostatic compression that does not permit the appropriate measurement of maximum recoverable strain.[20] The co-existence of two different crystal structures induces forming and breaking of P-P bonds at different stress levels, which is clearly displayed as two steps in the experimental stress-strain curve. The cyclic loading-unloading tests confirm that this double lattice collapse-expansion process occurs stably during ~10$^4$ loading-unloading cycles, indicating that a SrNi$_2$P$_2$ single crystal is truly pseudoelastic and structurally stable. The double lattice collapse-expansion mechanism shows stress-strain behaviors similar to that of conventional shape memory alloys, such as hysteresis and thermo-mechanical actuation, even though the structural changes involved are completely different. Its high yield strength (3.8±0.5GPa) and large maximum recoverable strain bring out the ultrahigh modulus of resilience (146±19MJ/m$^3$) a few orders of magnitude higher than that of most engineering materials. Note that SrNi$_2$P$_2$ exhibits the ThCr$_2$Si$_2$ structure, one of the most common, ternary intermetallic structures known, and many ThCr$_2$Si$_2$-type phases are believed to exhibit the lattice collapse-expansion process.[21,22] Therefore, our work suggests that a new class of pseudoelastic materials, consisting of a large number of different intermetallic ThCr$_2$Si$_2$-type phases, can be potentially discovered, leading to new research opportunities in the field of pseudoelasticity.

## 2. Results and Discussion

## 2.1 Co-existence of two different SrNi$_2$P$_2$ structures

SrNi$_2$P$_2$ has been known to exhibit polymorphism with two different crystal structures, the orthorhombic (O) superstructure (Figure 1a) and the tetragonal (T) structure (Figure 1b).[20,23,24] The O superstructure is the low temperature form that is present below 325K, whereas the T structure (the standard ThCr$_2$Si$_2$ structure) is present above 325K.[20] We grew a single crystal of SrNi$_2$P$_2$ (the inset of Figure 1c) using the solution growth method (See also Supporting Information S1). The first-order phase transition at around 325 K manifests itself clearly in thermodynamic and transport properties and can be seen in the hysteretic, temperature-dependent electrical resistance data (Figure 1c). The O superstructure resembles three laterally connected T structures, but with two different P-P distances ($d_{P-P,1}$ and $d_{P-P,2}$ in Figure 1a). Powder X-ray diffraction (XRD) on our SrNi$_2$P$_2$ single crystal, shows that the O superstructure and T structure co-exist (See also Supporting Information S3). By comparing the intensities of XRD peaks with those of calculated peaks for the individual structures, we estimated the volume fractions as O:T = 83.5%:16.5%. Thus, the O superstructure occupies most of the sample volume.

To search for direct evidence of the co-existence of these two structures, transmission electron microscopy (TEM) was performed along the [010] zone axis. The co-existence of the two phases was captured at an intermediate magnification (110,000X) (See also Supporting Information S4). It was possible to identify the Fast Fourier Transform (FFT) patterns of both structures in a single image. High-resolution TEM (HRTEM) was also performed to check for the existence of the two different atomic arrangements. The FFT patterns of these HRTEM images match the calculated diffraction patterns for the individual structures perfectly (Single Crystal 4.0 software, Crystal Maker Software Ltd) (Figures 1d-1i). In summary, the X-ray diffraction data and the TEM analysis confirmed that the O superstructure and the T structure coexist in our sample. Note that all TEM images here were taken before the electron-beam-induced transformation occurs (See also Supporting Information S5 and Supporting Movie 1).

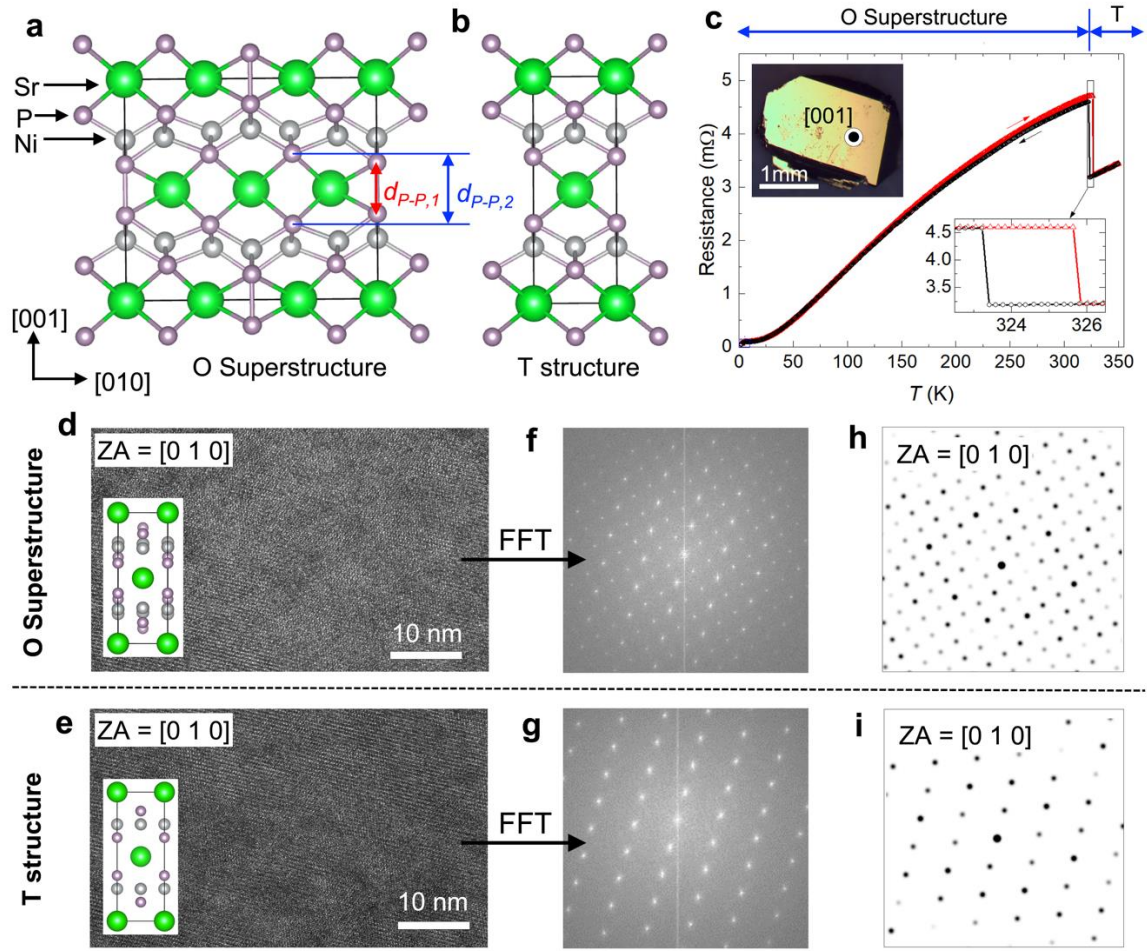

**Figure. 1 Structures and electrical properties of SrNi$_2$P$_2$.** (a) orthorhombic (O) superstructure based on experimental data (a = 3.951Å, b = 11.853Å, c = 10.432Å, the P-P distances are 2.45Å and 3.28Å) (b) tetragonal (T) structure based on experimental data (a = b = 3.948Å, c = 10.677Å, the P-P distance is 3.12Å). (c) the temperature-dependent electrical resistivity. The inset shows an optical microscope image of solution-grown single crystalline SrNi$_2$P$_2$. (d) HR-TEM image of the O superstructure with a [0 1 0] zone axis; the inset is the atomic structure of the O superstructure viewed along a [0 1 0] direction. (e) HR-TEM image of the T structure with a [0 1 0] zone axis;

the inset is the atomic structure of the T structure viewed along a [0 1 0] direction. (f) FFT pattern of (d). (g) FFT pattern of (e). (h) Simulated diffraction pattern of the O superstructure in Figure 1a along a [0 1 0] zone axis. (i) Simulated diffraction pattern of the T structure in Figure 1b along a [0 1 0] zone axis.

## 2.2 Uni-axial compression of SrNi$_2$P$_2$ micropillars along the c-axis

According to the low hydrostatic pressure (4 kbar) for lattice collapse in SrNi$_2$P$_2$, it was expected that SrNi$_2$P$_2$ could exhibit the lattice collapse along the c-axis at a relatively low uniaxial stress before the plasticity or fracture occurs.[23] Therefore, uniaxial compression tests on SrNi$_2$P$_2$ were performed along the c-axis. Interestingly, the strain-stress data for SrNi$_2$P$_2$ shows five distinct stages of deformation (Figures 2a and 2b) prior to fracture. First, the initial elastic deformation occurred (Stage I). Then, the first lattice collapse occurred at ~0.28GPa (Stage II). And then the second elastic deformation occurred at only slightly higher stress (Stage III). The larger second lattice collapse occurred at ~0.33GPa (Stage IV), followed by the large third linear elastic deformation (Stage V). Finally, brittle fracture occurred at the end of the Stage V elastic deformation. Overall, the stress-strain data clearly show the presence of two steps (double lattice collapse) in the loading curve as well as the large maximum recoverable strain, ~14.7%, which was also confirmed by the in-situ recorded video (Figure 2c and see also Supporting Movie 2). To determine the maximum recoverable strain precisely, we performed the three loading-unloading tests with the incremental maximum target load as suggested by Maaß and Derlet.[25] (See also Supporting Information S6) and two single loading tests. In case of the single loading test, the maximum recoverable strain was then obtained by subtracting the total plastic strain from the fracture strain. All five tests report that the average maximum recoverable strain is 14.1±0.4% (min: 13.5%, max: 14.7%).

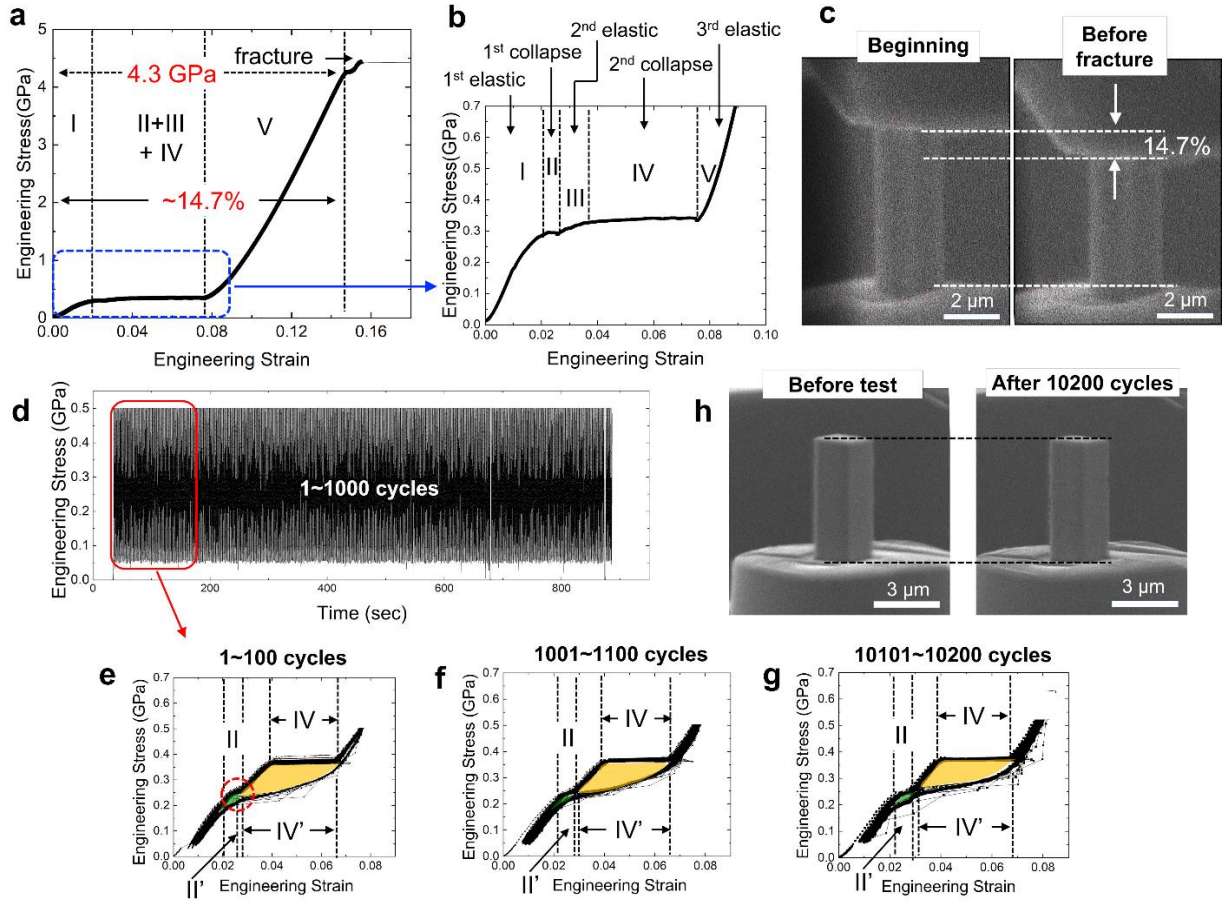

**Figure. 2 Pseudoelastic deformation of SrNi$_2$P$_2$.** (a) Experimental stress-strain curve of a [001]-oriented SrNi$_2$P$_2$ micropillar up to the fracture strength. (b) Magnified stress-strain curve in blue region in (a). (c) Snapshots of in-situ compression test. (d) Stress-time data of the first 1,000 cycle test. (e) Stress-strain data of the first 100 cycles. (f) Stress-strain data of 1,000-1,100 cycles. (g) Stress-strain data of 10,100-10,200 cycles. (h) SEM images before and after cyclic test.

To confirm that the double lattice collapse can be reversed when an applied load is relaxed, loading-unloading cyclic tests were performed up to 10,200 cycles (Figure 2d-2h and see also Supporting Movies 3). The cyclic stress-strain curves show that the five stages of deformation occur repeatedly in both loading and unloading curves (Figures 2e, 2f, and 2g). Thus, not only does the double lattice collapse occur during loading, but also the double lattice expansion occurs during unloading (Stages II' and IV' in Figures 2e, 2f, and 2g). The stress-strain curves of the first 100 cycles (Figure 2e) were compared with the 1000~1100 cycles (Figure 2f) and the last 100 cycles (Figure 2g) to confirm the stability of double lattice collapse-expansion process. We found tiny differences in the cyclic stress-strain data (red broken circle in Figure 2e), but these differences disappear nearly at the end of 100 cycles. Subtle differences may result from the initial tip-sample contact related to sample roughness and small misalignment. After 100 cycles, the results show negligible differences in stress-strain data up to 10,200 cycles, implying that the double lattice-collapse processes are fully reversible and repeatable. Also, we confirmed that almost no change in height was detected (Figure 2h), implying that there was no measurable permanent damage, such as the evolution of dislocation structures or microcracking. Therefore, the cyclic tests and SEM images prove that all five stages of deformation are recoverable and repeatable, indicating that $SrNi_2P_2$ is truly pseudoelastic and structurally stable. Note that the dislocation plasticity or shear-based deformation twinning could not be responsible for this first small strain burst (Stage II), which is recoverable, because there is no microstructural feature that can induce the backward motion of dislocations, and it is difficult to form the recoverable deformation twin in the complex atomic configuration of two-phase single crystal $SrNi_2P_2$.

The lattice collapse-expansion mechanism shows similar features to those of the martensite-austenite phase transformation of conventional shape memory alloys, although the structural changes involved are completely different. As a first order phase transition, an abrupt change in shape appears as a plateau in the stress-strain curve. The hysteresis loop exists in the loading-unloading stress-strain curve (Figures 2e, 2f, and 2g). The hysteresis can be present when the potential energy – strain curve has two minima, and a material is deformed under load-control. Our previous density functional theory (DFT) simulations on $CaFe_2As_2$ confirmed that the shape of the potential energy – strain curve is similar to that of conventional shape memory alloys.[17,26] In addition, the temperature dependence of the transition stress is similar. Our in-situ cryogenic micropillar compression shows that the structural transition occurs at a lower stress when the temperature decreases (Figure 3a).[27] As the temperature decreases, the two phases would become structurally similar, resulting in a structural transition at a lower stress.[28] In the case of shape memory alloys, the lattice parameter of the martensite phase becomes similar to that of the austenite phase as the temperature decreases. For $SrNi_2P_2$, the structure of the collapsed phase, which is denser, may be less sensitive to temperature while the open structure of the un-collapsed phase may reduce its lattice parameter more rapidly as the temperature decreases. So, the structure of the un-collapsed phase becomes more similar to that of the collapsed phase at a lower temperature, leading to the lattice collapse at a lower stress. As a consequence, the differences in the critical stress for the structural transition at different temperatures could cause differences in height under pre-stressed conditions, enabling the thermal actuation (Figure 3b). Therefore, all these results suggest that the theoretical framework, which is associated with the strain-dependent potential energy and the temperature effects, of conventional pseudoelastic and shape memory materials could be applied to describe the lattice collapse and expansion behavior of $SrNi_2P_2$.

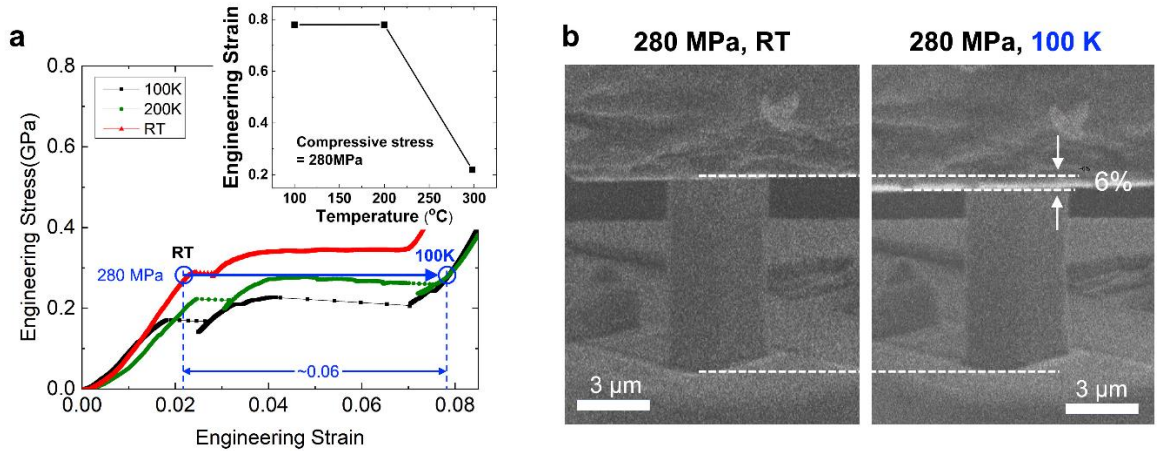

**Figure. 3 Temperature-dependent mechanical behavior of SrNi$_2$P$_2$.** (a) Stress-strain data at room temperature, 200K and 100K. The inset shows that structural transition occurs at lower stress when temperature decreases. (b) SEM image at constant stress (280 MPa) and different temperature (RT and 100K). The strain difference, 6%, is shown as the blue arrow in (a).

**2.3 Mechanism of Double Lattice Collapse**

To understand the atomic-scale process of lattice collapse in detail, we performed DFT calculations of uniaxial compression on the O superstructure and the T structure (Figure 4). In the undeformed state, the c-axis lattice parameter of the O superstructure (10.354Å) is shorter than that of the T structure (10.555 Å), and the collapsed tetragonal structure has an even shorter c-axis lattice parameter (9.111 Å). Our TEM results showed that SrNi$_2$P$_2$ is a two-phase single crystal because there is no separate set of diffraction pattern. This implies that the O superstructure and the T structure are coherently connected. Thus, the O superstructure region would be slightly under tension and the T structure region would be slightly under compression along the c-axis to meet the strain compatibility. Once the external compressive stress increases, the T structure experiences an even higher compression, and the O superstructure relaxes the residual compressive stress. Then, the T structure could be transformed into the O superstructure to relax the high compressive stress. The direct transformation from the T structure to its collapsed tetragonal structure would not occur because the lattice parameter of the collapsed tetragonal structure is smaller than that of the O superstructure. In addition, the small transformation strain of the first collapse implies that the first collapse is related to the phase transformation of the small volume fraction phase, which is the T structure as seen in XRD analysis (Supporting Information S3). The DFT simulation results show that the T structure undergoes the structural transition (the formation of P-P bonds) at a lower stress (372 MPa) than that of the O superstructure (634 MPa) (Figure 4a). Based on all experimental and simulation results, the small strain burst at a lower stress in Stage II could correspond to the phase transformation from the T structure to the O superstructure by forming P-P bonds partially (Figure 4b). Then, the micropillar, which is fully occupied by the O superstructure, collapses to the collapsed tetragonal structure under further compression (Figure 4b). The large strain burst in Stage IV corresponds to the lattice collapse of the entire sample volume.

The second collapse in Stage IV can be understood by the following bonding process. The calculated electronic band structure shows the bonding and antibonding states associated to the $3p_z$ orbitals of different P sites in the O superstructure (before collapse, 3% compression) and the collapsed tetragonal structure (after collapse, 9% compression) (Figure 4c). In the O superstructure, the antibonding state for $P_1$ is above the Fermi level suggesting the $P_1$-$P_1$ bonding, while the antibonding state for $P_2$ crosses the Fermi level, since the $P_2$-$P_2$ distance is too large for bond formation. Upon increasing uniaxial stress, the O superstructure transforms into the collapsed tetragonal structure where all P sites are equivalent, and the compression ratio abruptly increases to around 9%. The band structure shows that the antibonding state for all P sites is above the Fermi level, which indicates the bond formation throughout the whole structure after the collapse transition (Figure 4c). We note that, in this picture, the O superstructure is 1/3-collapsed compared to the T structure at zero pressure where the antibonding state for all P sites crosses the Fermi level (results not shown). Although DFT simulation has its own limitation because it deals with an energetically stable structure at 0K instead of the room temperature at which we perform the compression tests, our computational results are well in line with the experiments. Thus, the DFT simulations are indeed presenting a reliable framework to understand our results.

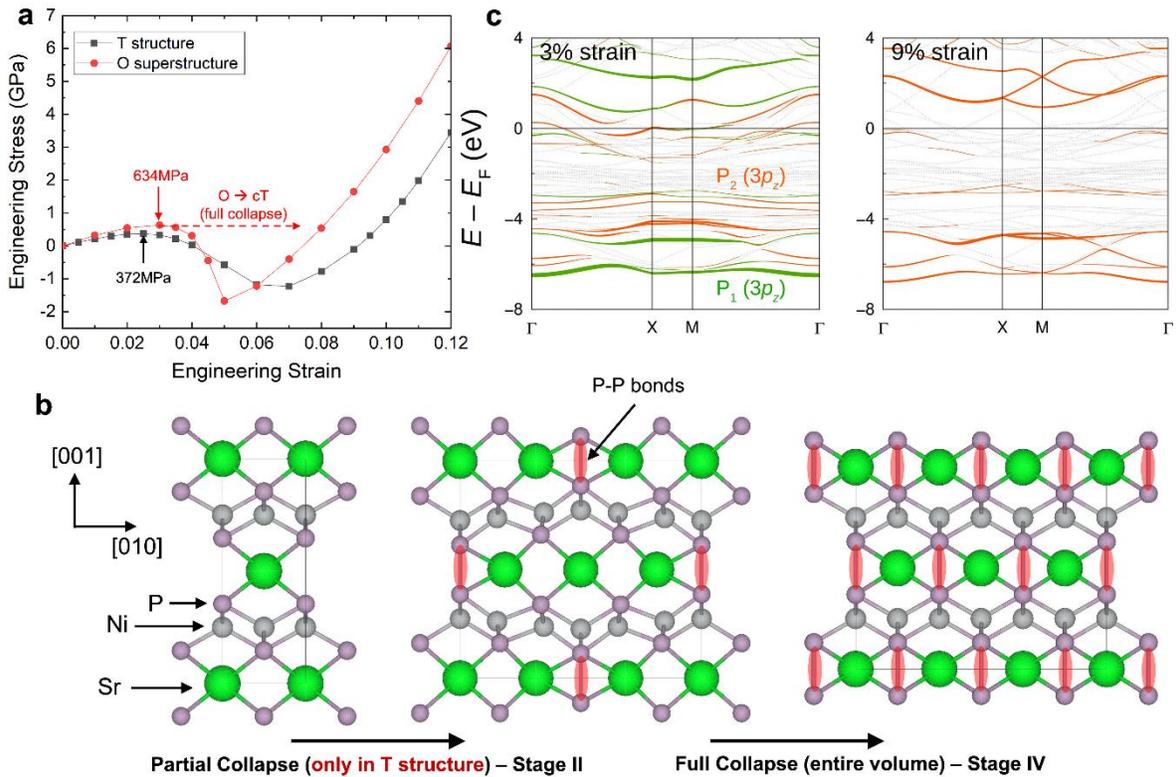

**Figure. 4 DFT analysis of uni-axial compression test.** (a) DFT stress-strain data of the T structure and the O superstructure. The broken red line represents the stress-strain path under the experiment-like condition (load control) for the lattice collapse in Stage IV, the phase transformation of the entire pillar from the O superstructure to the collapsed tetragonal structure.

(b) Schematic diagram of structural changes during the double lattice collapse. Note that the first collapse occurs only in the T structure, which occupies the small fraction of volume. (c) Band structure of the O superstructure at 3% strain (left) and that of the collapsed tetragonal phase at 9% strain (right).

## 2.4 Pseudoelastic Performance

The maximum recoverable strain of $SrNi_2P_2$ is 14.1±0.4%. This value is surprisingly high compared to any other crystalline materials (Figure 5a). Most bulk shape memory alloys and ceramics exhibit a much lower maximum recoverable strain (~up to 10%) with some exceptions.[29] Size reduction down to the nanometer scale often enhances the maximum recoverable strain. NiTi nanopillars with sub-200nm in diameter show an maximum recoverable strain of up to 15%.[30] $Ni_{55-x}Co_xFe_{18}Ga_{27}$ single-crystal fibers with 30-500μm in diameter exhibit a maximum recoverable strain of 15.2%.[31] In case of $ThCr_2Si_2$-structured intermetallic compounds, $LaRu_2P_2$ and $CaFe_2As_2$ exhibit the maximum recoverable strains of ~5% and ~11%.[17,19] Our recent studies on the hybrid structure, $CaKFe_4As_4$ gave values of 14~17%.[18] Note that the value for $SrNi_2P_2$ obtained in this study (using columns with dimensions of micrometers, not nanometers) exceeds or similar with the maximum recoverable strain of the state-of-the-art crystalline solids. The elastically compliant Sr-P layers (See also Supporting Information S7) and the double lattice collapse-expansion enables the large maximum recoverable strain of $SrNi_2P_2$. The large maximum recoverable strain and the relatively high yield strength allows $SrNi_2P_2$ to have an ultrahigh modulus of resilience, the absorption of mechanical work per unit volume until plastic yielding or fracture occurs (Figure 5b).

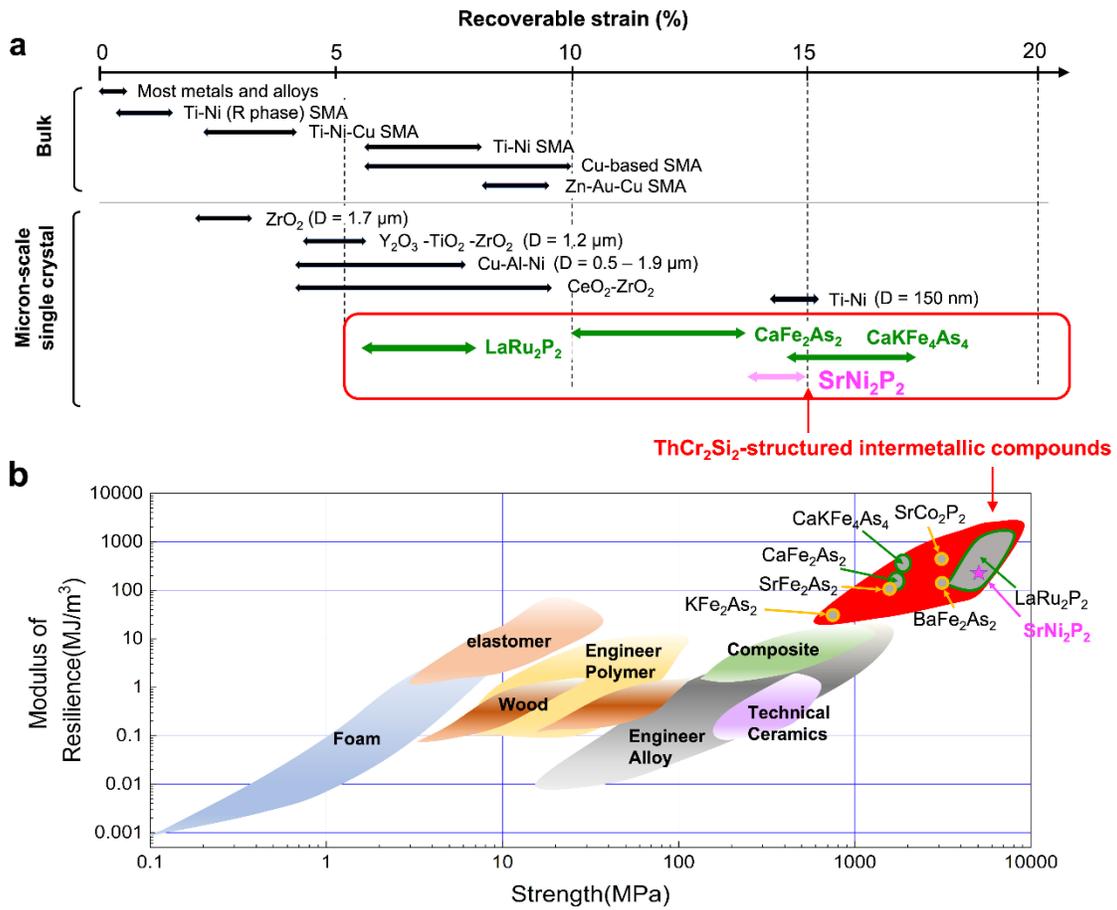

**Figure. 5 Pseudoelastic performance of SrNi$_2$P$_2$.** (a) Maximum recoverable strain of pseudoelastic materials at different length scales. ThCr$_2$Si$_2$-structured intermetallic compounds and similar structures show an exceptionally high maximum recoverable strain. (b) Revised Ashby chart of modulus of resilience as a function of yield strength. ThCr$_2$Si$_2$-structured intermetallic compounds and similar structures show the modulus of resilience, the orders of magnitude higher than those of most engineering materials.

The modified Ashby Chart (Figure 5b)[32] shows that the modulus of resilience SrNi$_2$P$_2$ is 146±19MJ/m$^3$, which is orders of magnitude higher than those of most engineering materials. Note that pseudoelasticity in SrNi$_2$P$_2$ is achieved through a simple structural transition process, forming and breaking bonds, which is almost identical to the bond stretching and contracting involved in linear elasticity. Although high cycle fatigue tests with over a million cycles could not be performed due to the limitation of our micromechanical testing capability, thus, an excellent fatigue resistance is expected because the uniform occurrence of forming and breaking bonds in the entire volume of material does not generate a significantly strain-incompatible phase, nor does it induce the motion of dislocations. In sum, SrNi$_2$P$_2$ could provide a superior protection from the repeated application of extremely high mechanical energy along with other ThCr$_2$Si$_2$-structured intermetallic compounds (Figure 5b).

## 3. Concluding Remarks

This work shows that SrNi$_2$P$_2$, one of ThCr$_2$Si$_2$-structured intermetallic compounds, exhibits a large maximum recoverable strain (14.1±0.4%%) via unique reversible structural transition, double lattice collapse-expansion. The forming and breaking of P-P bonds in the T structure and the O superstructure at two different stress states are the primary reasons for double lattice collapse-expansion. Due to the simple bonding-debonding process and the transition stress much lower than the yield strength, it is expected that SrNi$_2$P$_2$ exhibits the superior fatigue resistance. In addition, its high yield strength (3.8±0.5GPa) and large maximum recoverable strain bring out the ultrahigh modulus of resilience (146±19MJ/m$^3$), implying that SrNi$_2$P$_2$ can absorb and release the huge amount of mechanical energy. Note that $SrNi_2P_2$ is only one example among 2500 $ThCr_2Si_2$-structured intermetallic compounds that are believed to exhibit the pseudoelasticity via the lattice collapse-expansion mechanism.[16,21,22] Therefore, our work not only provides an insight into fundamental understanding of pseudoelasticity mechanism of $SrNi_2P_2$ but also suggests a potential discovery of $ThCr_2Si_2$-structured pseudoelastic materials that are fundamentally different from a conventional pseudoelastic and shape memory materials. This discovery will open a new research opportunity for the development of high maximum recoverable strain, high fatigue resistance, and strain-engineerable pseudoelastic materials.

ASSOCIATED CONTENT

**Supporting Information**

The following files are available free of charge.

- Materials and Method; Mechanical properties of several $ThCr_2Si_2$-structured intermetallic compounds; loading-unloading stress-strain data of three additional tests of $SrNi_2P_2$; volume fraction calculation of two co-existing structures in $SrNi_2P_2$; additional TEM images and e-beam induced transformation; DFT data of the gap between Si-P layers during compression. (PDF)

- Video S1: Movie of electron-beam induced phase transformation. (MP4)

- Video S2: Movie of in-situ micropillar compression. (MP4)

- Video S3: Movie of in-situ cyclic test. (MP4)

AUTHOR INFORMATION

**Corresponding Author:**

**Shuyang Xiao** - *Department of Materials Science and Engineering & Institute of Materials Science, 97 North Eagleville Road, Unit 3136, Storrs, CT 06269-3136, USA*; Email: shuyang.xiao@uconn.edu

**Author Contributions:**

S.X., P.C.C., and S.-W.L designed the experiments. S.X., G.S., J.M.M., S.R. and G.G.L. ran the experiments. V.B and R.V. ran the simulations, and V.B did the corresponding calculation. S.X., V.B., S.-W.L. interpreted the results. S.X. and S.-W.L. wrote the manuscript. V.B., M.A., R.V., and P.C.C. edited the manuscript.

**Notes**

The authors declare that they have no competing interests.

**Acknowledgement**

S.X, G.S. J.M.M. and S.-W.L. are supported by Early Career Faculty Grant from NASA's Space Technology Research Grants Program (NNX16AR60G), GE fellowship, and the UConn Interdisciplinary Multi-Investigator Materials Project. The FIB milling and TEM studies were performed using the facilities in the UConn/Thermo Fisher Scientific Center for Advanced Microscopy and Materials Analysis (CAMMA). G.G.L. and P.C.C. is supported by the U.S. Department of Energy, Office of Basic Energy Science, Division of Materials Sciences and Engineering. Ames Laboratory is operated for the U.S. Department of Energy by Iowa State University under Contract No. DE-AC02-07CH11358. G.G.L. is also supported by the Gordon and Betty Moore Foundation's EPiQS Initiative through Grant GBMF4411. R.V. acknowledges support by the Deutsche Forschungsgemeinschaft (DFG, German Research Foundation) for funding through TRR 288-422213477 (project A05 and B05) V.B. acknowledges the computational resources provided by the computer center of Goethe University Frankfurt.

Supplementary Materials for

# Pseudoelasticity of SrNi$_2$P$_2$ Micropillar via Double Lattice Collapse and Expansion

*Shuyang Xiao[1,*], Vladislav Borisov[2], Guilherme Gorgen-Lesseux[3], Sarshad Rommel[1], Gyuho Song[1], Jessica M. Maita[1], Mark Aindow[1], Roser Valentí[2], Paul C. Canfield[3], Seok-Woo Lee[1]*

*Corresponding author. Email: shuyang.xiao@uconn.edu

**This PDF file includes:**

Supplementary Text

Figs. S1 to S7

Movies S1 to S3

## S1. Materials and Methods

**Single crystal growth:** Single crystals of $SrNi_2P_2$ were grown from a quaternary melt with an initial composition of $Sr_{1.3}Ni_2P_{2.3}Sn_{16}$. High purity elements were placed into a 2ml, fritted alumina crucible set [1], sealed into an amorphous silica ampoule under a partial atmosphere of high purity Ar and placed in a vented box furnace [2]. The furnace was then heated up to 600°C over 4hrs, dwelled at 600°C for an additional 4hrs, heated up to 1150°C over 5hrs, dwelled at 1150°C for 24hrs, and then cool down to 650°C over 250hrs. At 650°C the ampoule was removed from the furnace and placed in a centrifuge for decanting of the excess melt from the $SrNi_2P_2$, crystalline phase. Plate-like single crystals with dimensions of ~ 3x3x0.2 $mm^3$ were common with some crystals reaching up to 6 times that volume.

**Microstructure characterization:** Powder X-ray diffraction measurements were carried out on a Rigaku miniflex diffractometer using Cu-K-alpha radiation. Electron-transparent specimens for TEM analysis were prepared in a Thermo Fisher Helios Nanolab 460F1 FIB-SEM using the lift-out method. During this process, these specimens were thinned gradually to ~100 nm with the extent of any ion beam damage being minimized by reducing the ion beam current iteratively to 24 pA. TEM imaging and diffraction experiments were performed on these specimens in a Thermo Fisher Talos F200X scanning transmission electron microscope operated at an accelerating voltage of 200 kV.

**Micropillar fabrication:** The Thermo Fisher Helios Nanolab 460F1 FIB machine was also used to fabricate square cross-section micropillars 2μm in width and 6μm in height. Gallium ion beam currents from 300 to 10pA were used from initial to final thinning under an operation voltage of 30kV. Because the typical thickness (~20nm) of FIB damage layer is much thinner than the width of micropillar (~2μm), FIB damage effects on mechanical data are expected to be negligible. Size effect on superelasticity could also be negligible due to the much smaller length scale of phase transformation (the size of unit cell ~ a few angstroms) compared with the length scale of micropillars (a few micrometers). [3]

**In-situ nanomechanical testing:** In-situ nanomechanical compression tests were performed under ultra-high vacuum condition ($<10^{-4}$Pa) at different temperatures and different mechanical parameters using NanoFlip™ (KLA., TN, USA), which is installed in a field-emission gun JEOL 6335F scanning electron microscope (JEOL, Japan). The nominal displacement rate of 10nm/s was used and corresponds the engineering strain rate of around $0.0016s^{-1}$. Low temperature tests were performed by using the customized cryostat and liquid helium. The details of the cryogenic system are available elsewhere [3]. Cyclic tests were performed with displacement rate of 5000nm/s. Instead of performing 10,000 cycles loading at one time, we split it into ten times and 1,000 cycles loading per time given the limited processing ability of our software. For all tests in this study, thermal equilibration between the tip and sample was maintained, so the thermal drift was always below 0.5nm/s. Recorded videos were used to visually improve the accuracy of our strain measurement. The average stiffness of the micropillar base was obtained using equation,

$$d_{raw\ (P=P_{max})} = d_{sample\ (P=P_{max})} + \frac{P_{max}}{k_{base}} \qquad (1)$$

, where $d_{raw\ (P=P_{max})}$ is the raw displacement at the maximum load (Pmax), $= d_{sample\ (P=P_{max})}$ is the sample displacement at the maximum load measured from a recorded video, and $k_{base}$ is the average stiffness of the micropillar base. Then, the average stiffness was applied to all displacement data to obtain the sample displacement values by

$$d_{raw} = d_{sample} + \frac{P_{max}}{k_{base}} \qquad (2)$$

Here, $d_{raw}$ and $d_{sample}$ correspond to their all values measured from experiments.

**Density Functional Theory calculation:** The equilibrium crystal structure and electronic properties of SrNi$_2$P$_2$ under compressive strain were studied using density functional theory (DFT) [4,5] and ultrasoft pseudopotentials [6,7] available in the Quantum Espresso code [8,9]. In order to obtain reliable estimates of the critical pressure, we used the PBEsol functional [10] for the exchange-correlation energy, which partially circumvents the over- and underbinding problems of the usual LDA and GGA functionals in DFT. The electronic wavefunctions were represented using plane waves with an energy cutoff of 160 Ry and the charge density energy cutoff was set to 640 Ry. For the (1×1×1) structure of SrNi$_2$P$_2$, the integration over the first Brillouin zone was done on the Γ-centered (10×10×10) k-mesh, while, in case of the (1×3×1) structure of SrNi$_2$P$_2$, we used the Γ-centered (6×2×2) k-mesh. Methfessel-Paxton smearing with a width of 0.02 Ry was used for the electronic occupations.

In the first step, the crystal structure is fully optimized, assuming a non-magnetic state, and the resulting lattice parameters and internal atomic positions are considered as a reference corresponding to zero strain. For the (1×1×1) phase, the optimized lattice parameters are $a_0 = b_0 = 3.896$ Å and $c_0 = 10.56$ Å and the P-P interlayer distance is equal to 3.087 Å. For the (1×3×1) phase, we find that $a_0 = 3.909$ Å and the $b_0/a_0$ ratio deviates from 1 by $1.1\times10^{-3}$, while the lattice parameter $c_0 = 10.35$ Å. Due to the presence of the (1×3×1) superstructure, there are two kinds of interlayer P-P distances, equal to 3.782 Å and 2.661 Å. Starting from these equilibrium structures, compressive strain with a value ε is simulated then by fixing the lattice parameter along the [001] direction at a value $c = c_0(1 - \varepsilon)$ and optimizing the remaining lattice vectors and all internal atomic positions using the BFGS quasi-Newton algorithm. In the new equilibrium state, all components of the stress tensor are zero except for the component along the [001] direction, which is used for determining the stress-strain relation for the studied system.

The structural collapse transition was identified by analyzing the electronic band structure calculated using the all-electron full-potential localized orbitals (FPLO) basis set code [11] within the generalized-gradient approximation. The orbital-resolved analysis of the band structure allowed to determine the energy position of the bonding and antibonding P-3$p_z$-derived molecular orbitals relative to the Fermi level. The onset of the collapsed phase is indicated by the energy of the antibonding orbitals crossing the Fermi level and leading to the reduced occupation of the antibonding states [12,13].

## S2. Uniaxial stress-strain data of ThCr$_2$Si$_2$-structured intermetallic compounds

In CaFe$_2$As$_2$ (Figure S1a) and LaRu$_2$P$_2$ (Figure S1b), the lattice collapse occurs during uniaxial compression. After phase transformation, collapsed phase keeps being deformed elastically until fracture, so there are three stages of deformation (Stages I, II, and III). For the rest four compounds, BaFe$_2$As$_2$ (Figure S1c), KFe$_2$As$_2$ (Figure S1d), SrFe$_2$As$_2$ (Figure S1e) and SrCo$_2$P$_2$ (Figure S1f), fracture occurs before reaching the required critical stress for phase transformation. In order to observe the lattice collapse under uniaxial compression, therefore, the critical stress of lattice collapse must be lower than fracture strength.

## S3. Volume fraction of each structure

Powder X-ray diffraction (XRD) on our SrNi$_2$P$_2$ single crystal, shows that both orthorhombic (O) superstructure and tetragonal (T) structures seem to exist together (Figure S2a). Most diffraction peaks belong to those of the O superstructure, but some small peaks correspond precisely to those of the T structure (three green arrows in Figure S2a). Thus, our sample appears to mainly consist of the O superstructure.

Volume fraction of each structure can be obtained by comparing the intensity of their characteristic peaks in XRD patterns. After careful examination to our experimental powder XRD patterns, we found the characteristic peaks are at 2θ = 16.16° with intensity of 677 for the T structure and 2θ = 16.71° with intensity of 2996 for the O superstructure (Figure S2b). We also use VESTA software to generate the XRD patterns of the T structure and the O superstructure obtained from experimental data [14]. From the computed XRD pattern of each structure, we confirmed the position of characteristic peaks that match experimental XRD patterns; 2θ = 16.3° with intensity of 18.35 (Figure S2c) and 2θ = 16.63° with intensity of 15.97 (Figure S2d) for the T structure and the O superstructure, respectively. In order to get the volume fraction of two structures, we need to divide the intensities of two structures obtained from experimental powder XRD pattern by the corresponding intensity from simulated XRD pattern. Then we can get the relative quantity of T structure, 677/18.34 = 36.9 and O superstructure, 2996/15.97 = 187.6. We found that the T structure occupies 16.4% (=36.9/(36.9+187.6)) of the volume fraction while the O superstructure occupies 83.6% (=187.6/(36.9+187.6) of the volume fraction. This result is consistent with our mechanical data and TEM data, both of which show that the O superstructure occupies much more volume than the T structure in our sample.

S4. Low magnification TEM images and FFT patterns

To confirm co-existence of the T structure and the O superstructure, the selected area diffraction pattern was obtained, and the dark field image was taken using the two small diffraction spots that belong only to the O superstructure (yellow circle in Figure S3a). The TEM image showed the uniform contrast over the large area, implying that the dark field imaging is not able to visualize the T structure clearly. This result would be related to the structural similarity between the T structure and the O superstructure, which differs only by the small displacement of P atoms.

However, co-existence of the T structure and the O superstructure in $SrNi_2P_2$ specimen could be observed in the intermediate magnification TEM image (110,000 times). The Fast Fourier Transform pattern of the top right region show the presence of additional small spots (green arrows in Figure S4) and corresponds to the diffraction pattern of the O superstructure with a [010] zone axis. The FFT pattern of the bottom left region shows no small spots and corresponds to the diffraction pattern of T structure with a [010] zone axis. Therefore, the T structure and the O superstructure exist together in our sample.

**S5. Electron-beam induced phase transformation**

Videos of live FFT image were recorded in TEM (Thermo Fisher, USA). From the videos (See also Supplementary Movie 1), we found that the FFT patterns starts with both the small and the large spots, which corresponds to the O superstructure, but the smaller spots, which belong only to the O superstructure, disappear over time (Figure S5). We strongly believe that this phase transformation is due to the electron-beam-induced heating at the probed region. Temperature could increase above 325K, at which the structural transition occurs from the O superstructure to the T structure. At the highest magnification of 1.05 million times with 200kV accelerated electron voltage, it takes approximately 1min for the structure to transform. At 820,000 times magnification, it takes about 3mins for the structure to transform. At 650,000 times magnification, it takes around 4.5mins for the structure to transform (Figure S5). Presumably, the relationship of magnification and transformation time seem to be linear, and it should take at least 10 minutes for the structure to transform at low magnification of 110,000 times (Figure S6), which is much more than few seconds we used to take images in the actual experiment (Figures 2d, 2e, and S4). In this case, we can confirm that the T structure we captured at our TEM images is present with O superstructure instead of being transformed from O superstructure due to electron beam heating.

**S6. Measurement of the maximum recoverable strain**

Three loading-unloading tests with the incremental maximum target load as suggested by Maaß and Derlet[15] and two single loading tests were performed to determine the maximum recoverable strain. For loading-unloading tests, we gradually increased the target stress from a relatively large stress (~3GPa) until fracture occurs. In this case, we defined the maximum recoverable strain at which the first microplasticity event occurs in the final loading cycle (blue curve in Figure S6c-S6e), where the fracture occurs. In case of the single loading test, several micro-plasticity events were observed in Stage V. Thus, the maximum recoverable strain was then obtained by subtracting the total plastic strain from the fracture strain by using the average Young's modulus of Stage V that was obtained from the final loading curve of three loading-unloading tests. The average slope of Stage V curve was measured by connecting the start and the end of the elastic portion of Stage V curve (Figure S6). The use of the final loading curve for the measurement of the average Young's modulus of Stage V is beneficial because there is no microplasticity in the middle of Stage V. We think that the sources of microplasticity were exhausted in the prior loadings (black and red curve in Figure R6c-R6e), so the last loading curve does not show any stochastic microplasticity events until a sample exhibits macroscopic yielding right before fracture occurs. That's why the final loading curve of loading-unloading test has the smooth (i.e., purely elastic) Stage V curve. All five tests report that the average maximum recoverable strain is 14.1±0.4% (min: 13.5%, max: 14.7%) (Table S1).

## S7. The compliance of the gap between atomic layers

As FIG. 3A shown, Stage V contributes most significantly to the total elastic strain, which is the deformation of collapsed tetragonal structure. Note that due to the bonds twist, our Density Function Theory data shows that O superstructure has two different Sr-P and Ni-P distances in 1/3 part of superstructure along b axis solely which means the atoms of one layer are not in the same plane at 0 strain (Figure S7a). However, after the critical strain, which is 0.05 in our simulation, collapsed tetragonal structure has only a single Sr-P distance and a single Ni-P distance (Figure S7b). To examine the compliance of gap between atomic layers, we only need to discuss the contribution of layer compliances in collapsed tetragonal structure. After the collapse, the distance between Sr-P layers shrinks from 1.34Å to 1.19Å and distance of P-Ni layers decreases from 1.12Å to 1.09Å with increasing strain from 0.05 to the limit (Figure S7c). The data of change ratio, which is the ratio between the change of the distance of each layer and the change of the c lattice parameter, shows that the change of the distance of Sr-P layers accounts for 0.82-1.18 of the total strain while the change of the distance of Ni-P layers only contributes -0.15-0.18 to the total strain. This result shows that Sr-P layer gap contributes most to the total elastic strain (Figure S7d).

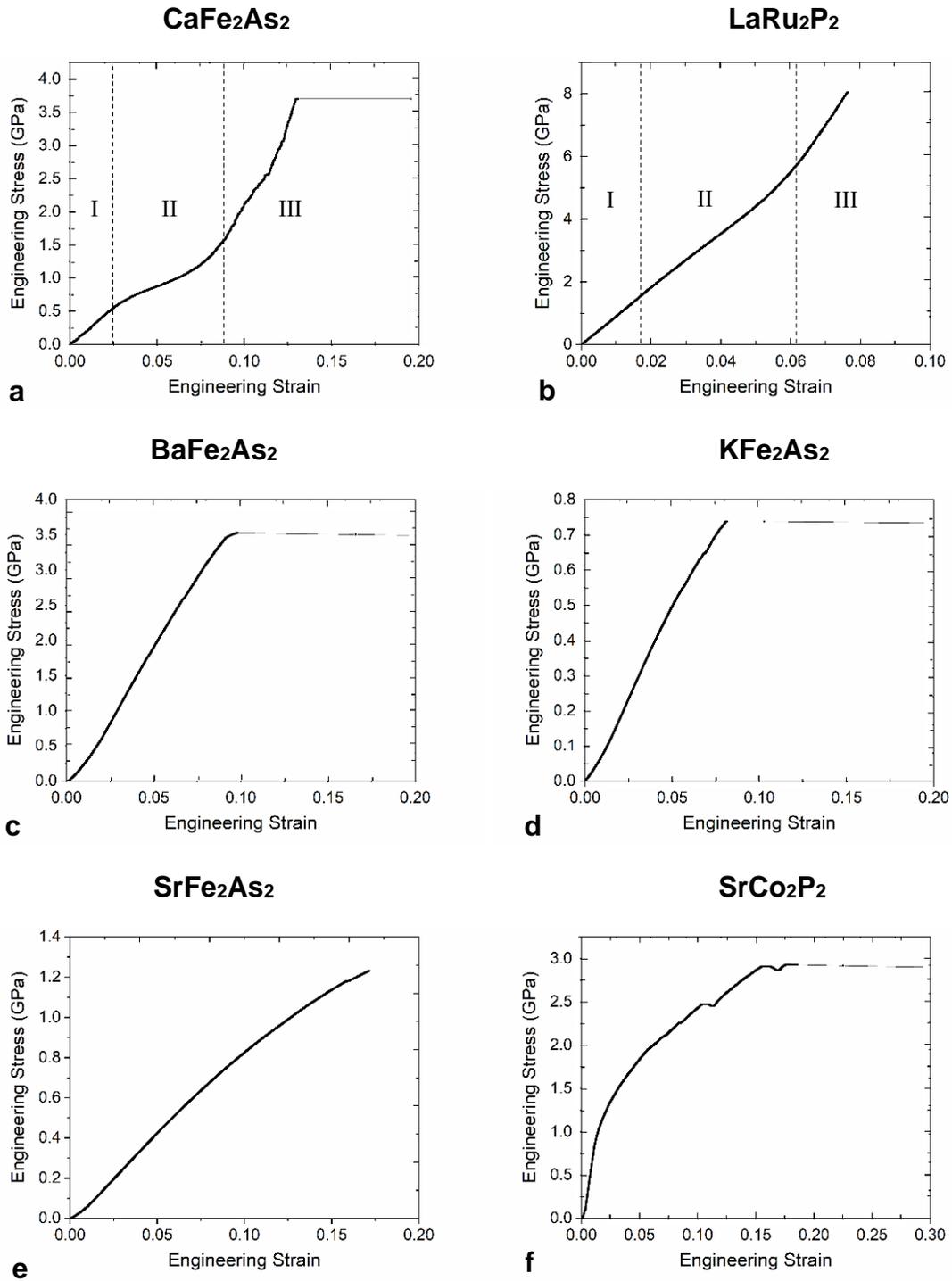

**Figure S1. Experimental stress-strain data.** (a) $CaFe_2As_2$ [16]. (b) $LaRu_2P_2$ [17]. (c) $BaFe_2As_2$. (d) $KFe_2As_2$. (e) $SrFe_2As_2$. (f) $SrCo_2P_2$.

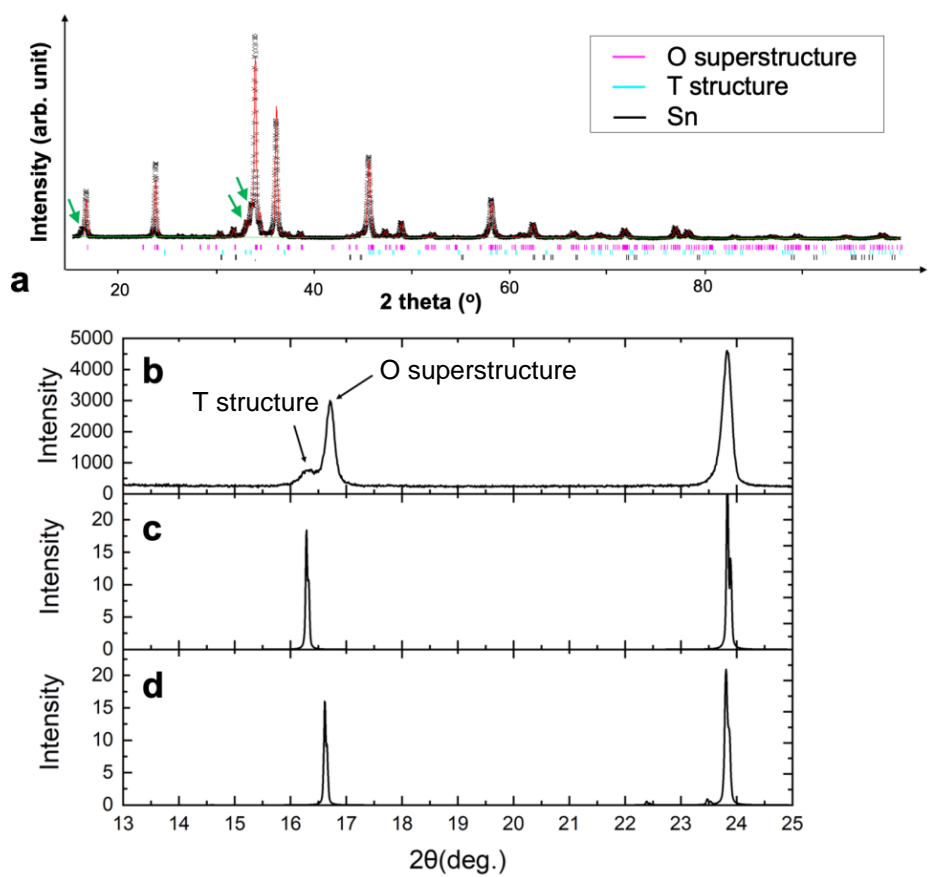

**Figure S2. Experimental and simulated X-ray diffraction (XRD) data.** (a) the experimental XRD data of SrNi$_2$P$_2$ (b) the experimental XRD data at the low $2\theta$ range. (c) the computed XRD data of the tetragonal structure at the low $2\theta$ range. (d) the computed XRD data of the orthorhombic superstructure at the low $2\theta$ range.

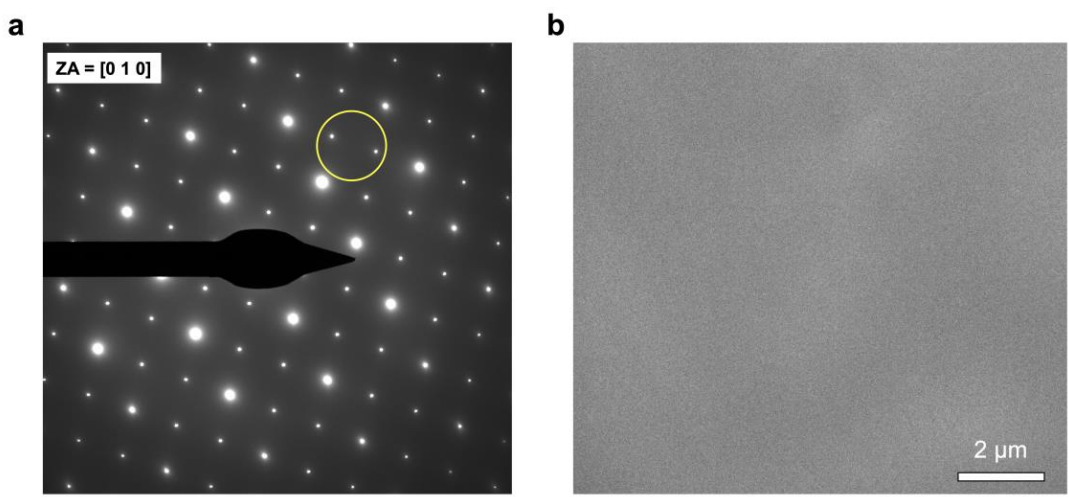

**Figure S3. Low Magnification Transmission Electron Microscope Image of SrNi$_2$P$_2$.** (a) the selected area diffraction pattern for the [0 1 0] zone axis. The aperture was located at the two small diffraction spots (the yellow circle) to obtain the dark-field image. (b) the dark-field TEM image obtained using the two small diffraction spots in (a).

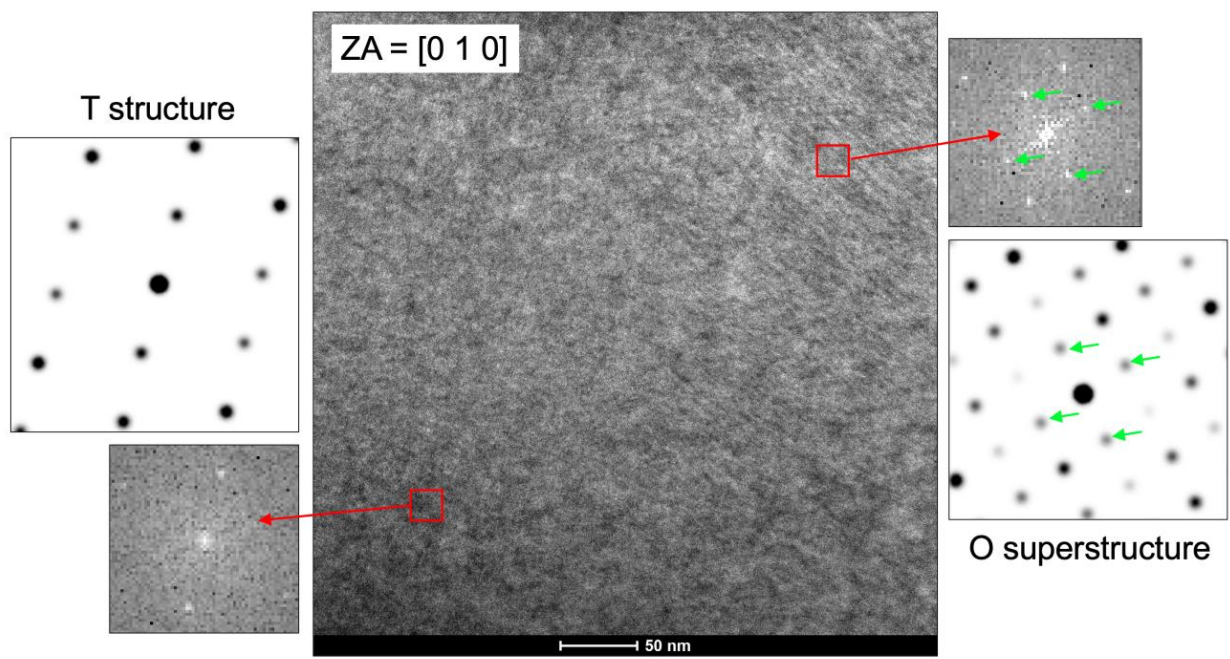

**Figure S4. Transmission Electron Microscope Image of SrNi$_2$P$_2$.** The small two insets correspond to the Fast Fourier Transform patterns of red-boxed regions, and the large two insets correspond to the simulated diffraction pattern of experimentally-obtained structures (Figures 1a and 1b).

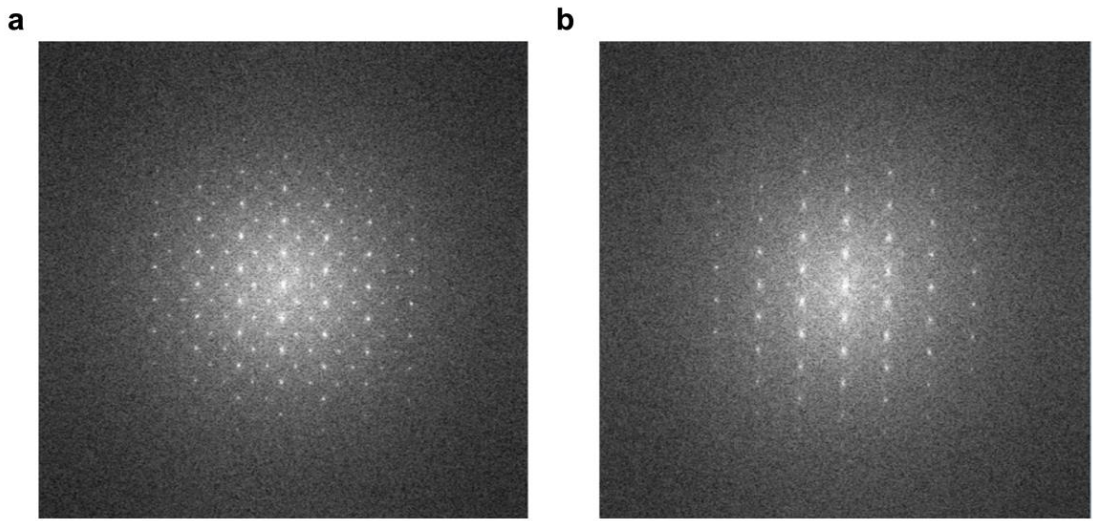

**Figure S5. FFT images of SrNi$_2$P$_2$ at 650000X magnification under 200kV electron beam.** (a) orthorhombic superstructure with small and large spots. (b) tetragonal structure with only large spots after 4.5 min electron beam exposure

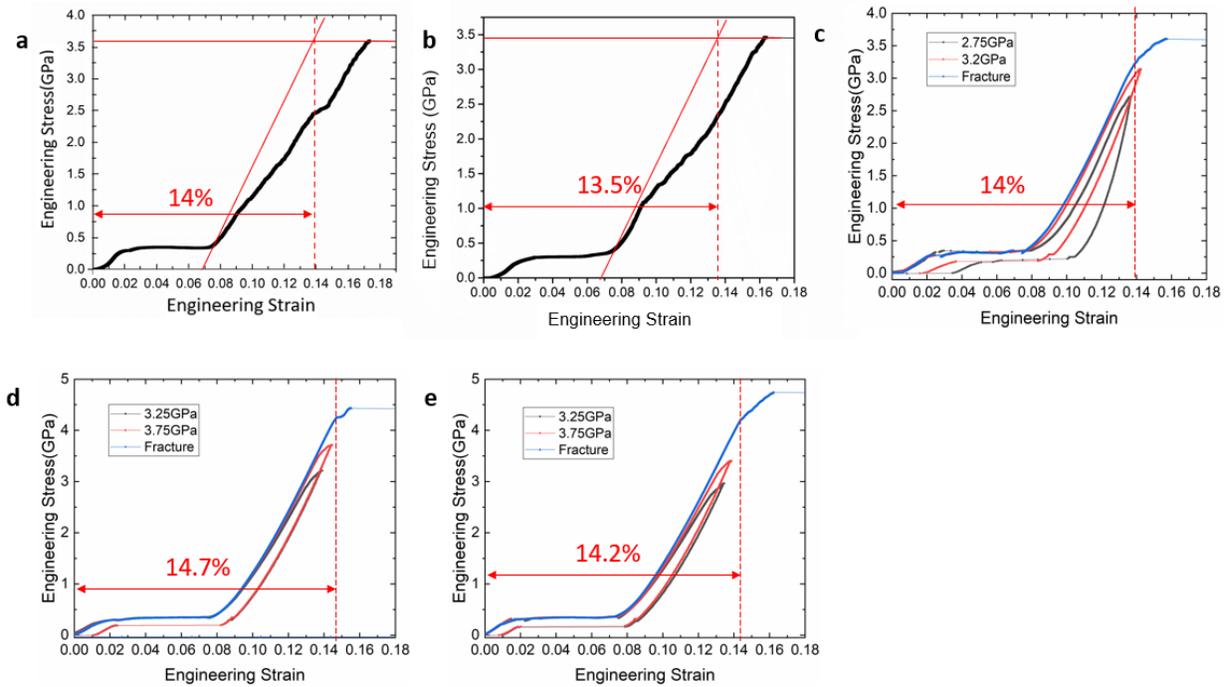

**Figure S6. Five tests to determine the maximum recoverable strain.** (a-b) Two single loading tests. (c-e) Three loading-unloading tests with incremental target load.

| Test | Test 1 | Test 2 | Test 3 | Test 4 | Test 5 | Average |
|---|---|---|---|---|---|---|
| **Maximum recoverable strain** | 14% | 13.5% | 14% | 14.7% | 14.2% | 14.1±0.4% |

**Table S1.** Summary of maximum recoverable strain for all 5 tests.

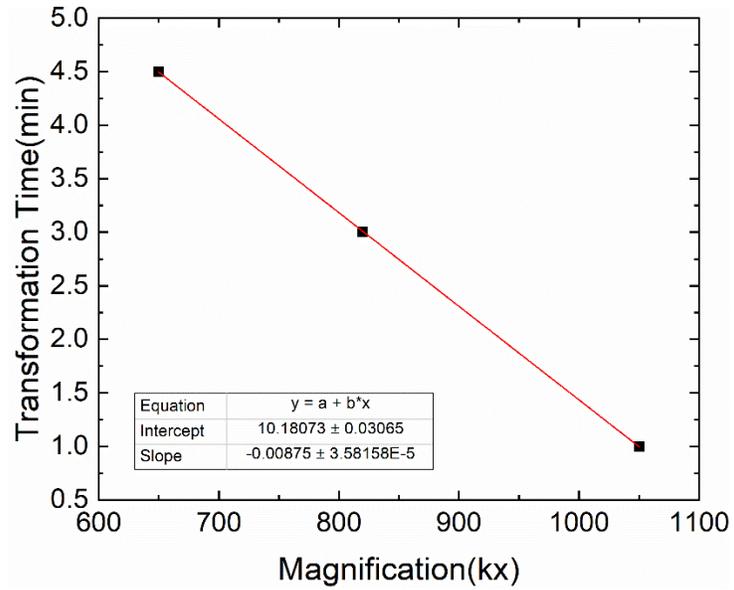

**Figure S6.** Estimated relationship between magnification and transformation time under transmission electron microscope.

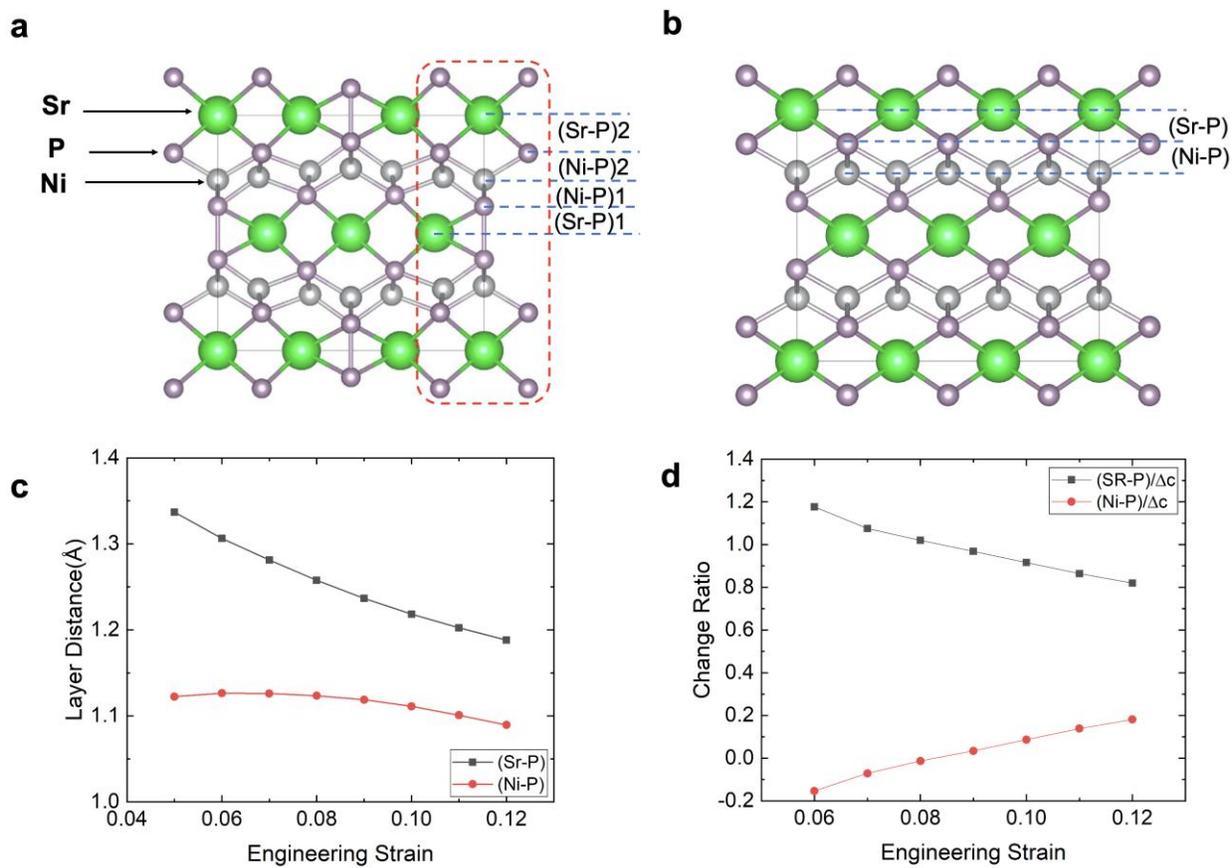

**Figure S7. Density Function Theory (DFT) data of contribution of different layers to elastic strain.** (a) orthorhombic superstructure at 0 strain. (b) collapsed tetragonal at 0.05 strain. (c) Layer distance change in collapsed tetragonal structure. (d) Change ratio of each layer in collapsed tetragonal structure.

**Movie 1.** E-beam induced transition in transmission electron microscopy at 650,000X magnification

**Movie 2.** Quasi-static uniaxial compression test

**Movie 3.** Cyclic test (the first 100 cycles).